\documentclass[11pt]{article}
\usepackage{graphicx}

\newcommand{\BABARPubYear}    {01}

\newcommand{\BABARProcNumber} {82}
\newcommand{\SLACPubNumber} {9034}
\def\lbabar{\mbox{{\large\sl B}\hspace{-0.4em} {\normalsize\sl A}\hspace{-0.03em}{\large\sl B}\hspace{-0.4em} {\normalsize\sl A\hspace{-0.02em}R}}}
\usepackage{relsize}
\def\babar{\mbox{\slshape B\kern-0.1em{\smaller A}\kern-0.1em
    B\kern-0.1em{\smaller A\kern-0.2em R}}}

\def\Kbar  {\kern 0.2em\overline{\kern -0.2em K}{}}

\def\Kzb   {\ensuremath{\Kbar^0}}
\def\KzKzb {\ensuremath{K^0 \kern -0.16em \Kzb}}

\def\Dbar  {\kern 0.2em\overline{\kern -0.2em D}{}}

\def\Dzb   {\ensuremath{\Dbar^0}}
\def\DzDzb {\ensuremath{D^0 {\kern -0.16em \Dzb}}}

\def\Bbar  {\kern 0.18em\overline{\kern -0.18em B}{}}

\def\Bzb   {\ensuremath{\Bbar^0}}

\def\BzBzb {\ensuremath{B^0 {\kern -0.16em \Bzb}}}

\mathchardef\Upsilon="7107
\def\Y#1S{\ensuremath{\Upsilon{(#1S)}}}% no space before {...}!

\mathchardef\Deltares="7101
\mathchardef\Xi="7104
\mathchardef\Lambda="7103
\mathchardef\Sigma="7106
\mathchardef\Omega="710A
\def\Deltabar   {\kern 0.25em\overline{\kern -0.25em \Deltares}{}}
\def\Lbar {\kern 0.2em\overline{\kern -0.2em\Lambda\kern 0.05em}\kern-0.05em{}}
\def\Sigbar{\kern 0.2em\overline{\kern -0.2em \Sigma}{}}
\def\Xibar{\kern 0.2em\overline{\kern -0.2em \Xi}{}}
\def\Obar{\kern 0.2em\overline{\kern -0.2em \Omega}{}}
\def\Nbar{\kern 0.2em\overline{\kern -0.2em N}{}}
\def\Xbar{\kern 0.2em\overline{\kern -0.2em X}{}}

\def\ev   {\ensuremath{\rm \,e\kern -0.08em V}}
\def\kev  {\ensuremath{\rm \,ke\kern -0.08em V}} 
\def\mev  {\ensuremath{\rm \,Me\kern -0.08em V}} 
\def\gev  {\ensuremath{\rm \,Ge\kern -0.08em V}} 
\def\gevc {\ensuremath{{\rm \,Ge\kern -0.08em V\!/}c}} 
\def\tev  {\ensuremath{\rm \,Te\kern -0.08em V}}
\def\mevc {\ensuremath{{\rm \,Me\kern -0.08em V\!/}c}} 
\def\gevcc{\ensuremath{{\rm \,Ge\kern -0.08em V\!/}c^2}} 
\def\mevcc{\ensuremath{{\rm \,Me\kern -0.08em V\!/}c^2}}

\def\mus  {\ensuremath{\rm \,\mus}}

\def\mus        {\ensuremath{\,\mu{\rm s}}}    %% microsecond
\def\gsim{{~\raise.15em\hbox{$>$}\kern-.85em
          \lower.35em\hbox{$\sim$}~}}
\def\lsim{{~\raise.15em\hbox{$<$}\kern-.85em
          \lower.35em\hbox{$\sim$}~}}

\def\to                 {\ensuremath{\rightarrow}}

\def\pep2{PEP-II}

\providecommand{\eqref}[1]{Eq.~(\ref{eq:#1})}

\def\jetset74   {\mbox{\tt Jetset \hspace{-0.5em}7.\hspace{-0.2em}4}}

\begin{document}
{\pagestyle{empty}

\begin{flushright}
SLAC-PUB-\SLACPubNumber \\
\babar-PROC-\BABARPubYear/\BABARProcNumber \\
%\babar-PUB-\BABARPubYear/\BABARPubNumber \\
%hep-ex/\LANLNumber \\
%October, 2001 \\
\end{flushright}

\par\vskip 4cm

% Title of the paper
\begin{center}
\Large \bf Three body decays of $D^0$ and $D_S$ mesons
\end{center}
\bigskip

\begin{center}
\large 
Antimo Palano\\
INFN and University of Bari, Italy\\
(for the \lbabar\ Collaboration)
\end{center}
\bigskip \bigskip

% Abstract
\begin{center}
\large \bf Abstract

\end{center}

Results are presented on the study of three body decays of 
$D^0 \to K^0_S h^+ h^-$, where $h=\pi/K$ and $D_S^\pm \to K^0_S K^0_S \pi^\pm$. The
data have been collected by the BaBar experiment at SLAC and are 
extracted from continuum $e^+ e^-$ annihilations
at the $\Upsilon(4S)$ energy.
\vfill
\begin{center}
Presented at Hadron 2001, IX International Conference on Hadron Spectroscopy \\
August 25--September 1, 2001, Protvino, Russia
\end{center}

\vspace{1.0cm}
\begin{center}
{\em Stanford Linear Accelerator Center, Stanford University, 
Stanford, CA 94309} \\ \vspace{0.1cm}\hrule\vspace{0.1cm}
Work supported in part by Department of Energy contract DE-AC03-76SF00515.
\end{center}
\section{Introduction}

New generation experiments are providing large data 
sets for charm physics with statistics which supersede most 
previous 
measurements.
The Dalitz plot analyses of 3-body charm decays have been performed in the
past but these new large and clean samples will allow high precision 
measurements that were never before possible.

The Dalitz plot analysis of three-body decays is a relatively new technique
in development for charm physics studies. 
This method of analysis is the
most complete way of analyzing the data since 
it allows measurement of both decay amplitudes and phases.
The final state is the result of the interference of all 
intermediate states.
The significant results provided by these studies are:
\begin{itemize}
\item{} Accurate measurements of branching fractions.
\item{} A study of Final State Interactions.
\item{} A study of CP violation in rates and decay amplitudes.
\item{} New input to several old unsolved problems in light meson 
spectroscopy, in particular to the scalar mesons puzzle.
\end{itemize}

Factorization models assume the weak decay amplitudes to be real.
The fact that the observed amplitudes have a relative complex phase
is a consequence of final state interaction. 

CP violation is expected to be small in charm decays 
($\approx 10^{-3}$)~\cite{cp}.
Two amplitudes with different phases are needed:
$$A e^{i\delta_A} + B e^{i\delta_B}$$
In singly Cabibbo-suppressed decays penguin terms may provide a weak phase,
while Final State Interactions provide a strong phase shift.
Under CP the weak phases change sign but the strong ones do not.
Any difference between $D$ and $\bar D$ in the Dalitz plot  
would be evidence for CP violation.

Throughout this paper charge conjugate modes, where not explicit, are implied.

\section{The BaBar Experiment}

The PEPII Collider is an asymmetric storage ring which 
collides 9 GeV electrons with 3.1 GeV positrons with a 
peak luminosity of 4.2$\times 10^{33}$ $cm^{-2} s^{-1}$.
The $\Upsilon(4S)$ resonance is produced with $\beta \gamma=0.56$ in 
the laboratory frame at zero crossing angle. Details on the layout of the
apparatus, trigger conditions and data processing can be found in previous
publications~\cite{babar1}.

The $\Upsilon(4S)$ resonance sits on a large continuum background 
with a contribution from $e^+ e^- \to c \bar c$ of 1.30 nb.
The power of BaBar for studying charm physics is based on:

\begin{itemize}
\item{Relatively small combinatorics because of $e^+ e^-$ interactions.}
\item{Good vertexing.}
\item{Good Particle Identification.}
\item{Detection of all possible final states, with charged tracks and $\gamma$'s.}
\item{Very high statistics.}
\end{itemize}

The BaBar experiment is continuously collecting data.
This work is based 
on the data taken during 1999/2000 and corresponds to an integrated
luminosity of $ \approx 23 fb^{-1}$ unless otherwise specified.

\section{Study of $D^+_S \to K^0_S K^0_S \pi^+$}
The resonance $f_J(1710)$ has been measured with spin 0 or 2 in different 
experiments. It has been a 
candidate for being the lowest lying scalar or tensor glueball. 
It has been observed 
in $J/\psi$ decay, central production and $\gamma \gamma$ 
collisions~\cite{theta}. Details on the selection criteria used to isolate 
the $K^0_S K^0_S \pi^+$ final state can be found in ref.~\cite{dep}.
The $K^0_S K^0_S \pi^+$ mass spectrum (relative to an integrated 
luminosity of 18.4 $fb^{-1}$) is plotted in fig.~\ref{fig:fig1} and 
shows signals from $D^+$ and $D_S^+$.
\begin{figure}[htp]
\begin{center}
\includegraphics[width=7cm,height=7cm]{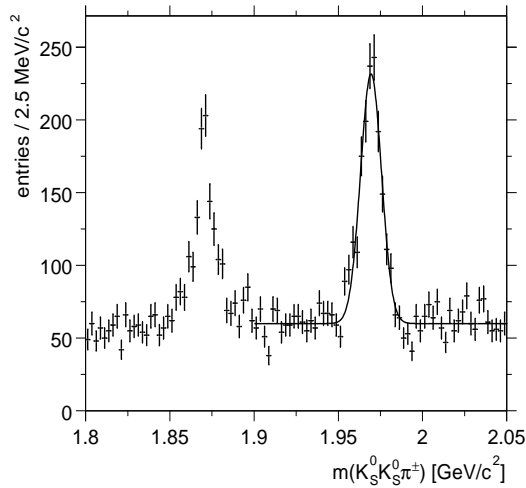}
\caption{$K^0_S K^0_S \pi^+$ mass spectrum.}
\label{fig:fig1}
\end{center}
\end{figure}
\begin{figure}[htp]
\begin{center}
\begin{tabular}{cc}
\includegraphics[width=7cm]{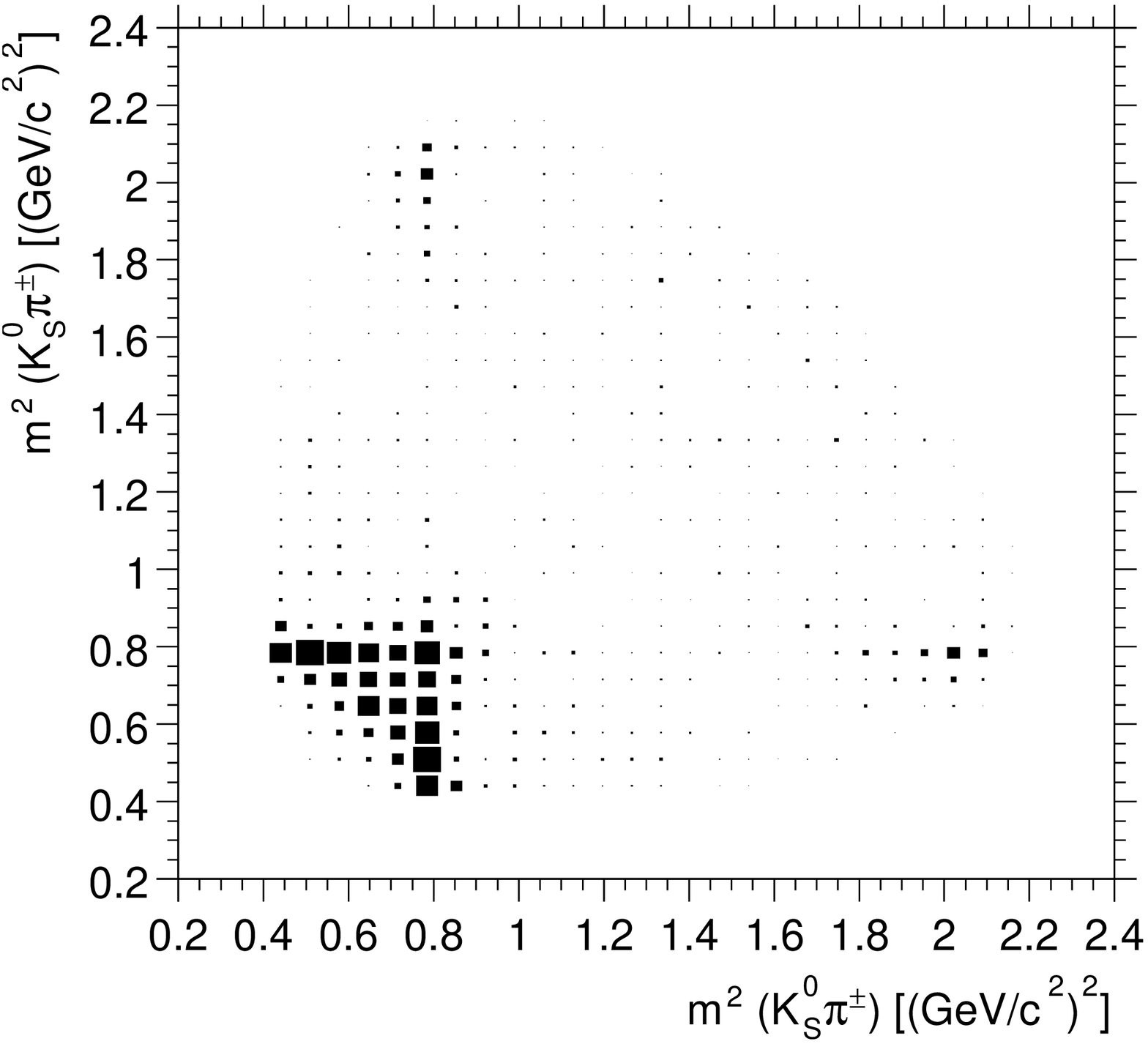} & \includegraphics[width=7cm]{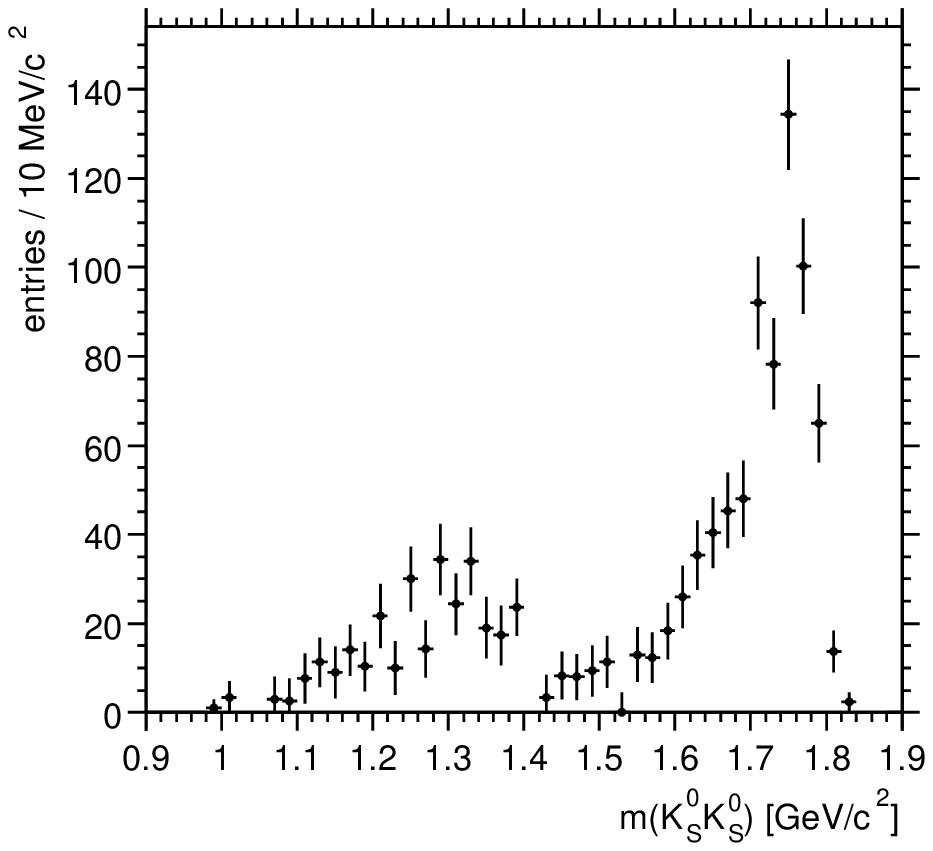}
\end{tabular}
\caption{Background subtracted Dalitz plot and $K^0_S K^0_S$ projection for
$D^+_S \to K^0_S K^0_S \pi^+$.}
\label{fig:fig2}
\end{center}
\end{figure}
The background subtracted Dalitz plot and its $K^0_S K^0_S$ 
projection 
are shown in fig.~\ref{fig:fig2}. Strong accumulations in the
region of the $K^*(892)$ can be seen. A preliminary Dalitz plot
analysis shows that the data cannot be described with $K^*(892)$ alone but
require the presence of the $D^+_S \to f_J(1710) \pi^+$ final state.
\section{Selection of $D^0 \to K^0_S h^+ h^-$}

$D^0$'s are required to come from a $D^{*+}$ decay:
$$D^{*+} \to \begin{array}[t]{l} D^0 \pi^+ \\ \to \bar K^0 \pi^+ \pi^- \end{array}$$
$$D^{*-} \to \begin{array}[t]{l} \bar D^0 \pi^- \\ \to K^0 \pi^+ \pi^- \end{array}$$
Therefore the charge of the slow $\pi$ gives the flavour of 
the $D^0$ and that of the $K^0$ (apart from the contribution from DCSD, 
which is expected to be $\approx 10^{-4}$).

The selection of the channel starts with the reconstruction of
the $K^0_S$ and $D^0$ vertices.
The slow $\pi$ are refitted using the beam spot constraint to 
improve the resolution,
(beam spot size:
$\sigma_x= 0.15 mm, \sigma_y= 0.05 mm, \sigma_z=8 mm$).
The center of mass momentum of the $D^0$ ($p^*$) has been required to be 
greater then 2.2 GeV/c.

The mass difference:
$$\Delta m = m(K^0 \pi^+ \pi^- \pi_s) - m(K^0 \pi^+ \pi^-)$$
where the slow pion $\pi_s$ has a momentum below 0.6 GeV/c is plotted in 
fig.~\ref{fig:fig3}. Fig.~\ref{fig:fig3}a) and fig.~\ref{fig:fig3}b) show respectively the   
$\Delta m$ distribution before and after having 
required a 2.5 $\sigma$ cut around the $D^0$ mass.
\begin{figure} [htp]
\begin{center}
\includegraphics[width=12.0cm,height=7.0cm]{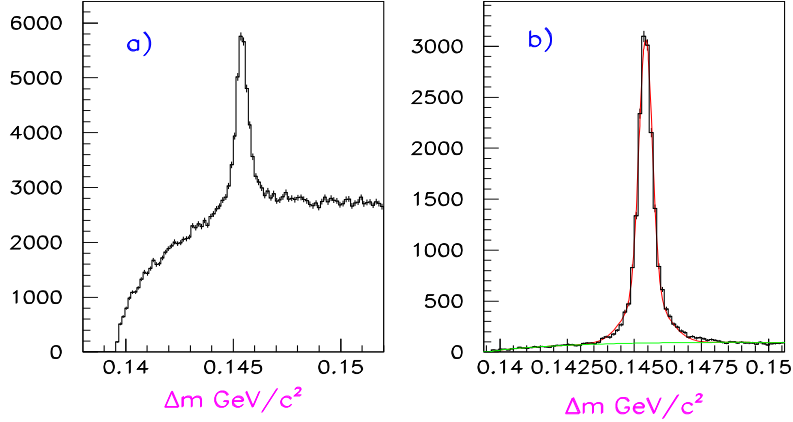}
\caption{$\Delta m$ distributions for $K^0_S \pi^+ \pi^-$ a) before and b)
after a $D^0$ mass cut at $2.5 \sigma$.}
\label{fig:fig3}
\end{center}
\end{figure}
\begin{figure} [htp]
\begin{center}
\includegraphics[width=7.0cm,height=7.0cm]{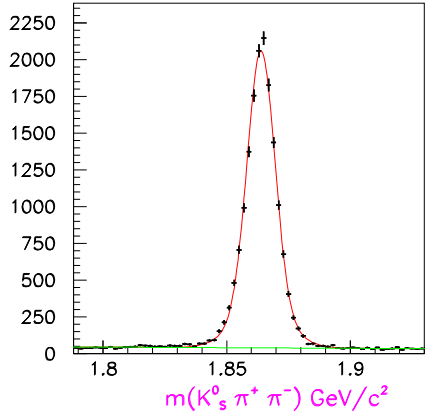}
\caption{$K^0_S \pi^+ \pi^-$ mass distribution after a 
$D^{*+}$ cut at $2\sigma$.}
\label{fig:fig4}
\end{center}
\end{figure}
Fitting the $\Delta m$ width using a single Gaussian and a threshold 
function $(m-m_{th})^\alpha e^{-\beta m - \gamma m^2}$,  we obtain
$\sigma = 326 \pm 10 \quad keV/c^2$.
Performing  now a 2.0 $\sigma$ cut on $\Delta m$  we obtain the 
$m(K^0_s \pi^+ \pi^-)$ shown in fig.~\ref{fig:fig4}.
The $D^0$ width obtained fitting only one Gaussian and a linear background is:
$\sigma = 6.3 \pm 0.1 \quad MeV/c^2$.

The Dalitz plot of $D^0 \to K^0_S \pi^+ \pi^-$, obtained by selecting events 
within 2.5 $\sigma$ of the $D^0$ mass, is shown in fig.~\ref{fig:fig5} 
(15753 events) and its projections are shown in fig.~\ref{fig:fig6}. 
The background fraction, estimated to be 4.1 \%, has not been subtracted. 
\begin{figure}[htp]
\begin{center}
\includegraphics[width=9.0cm,height=9.0cm]{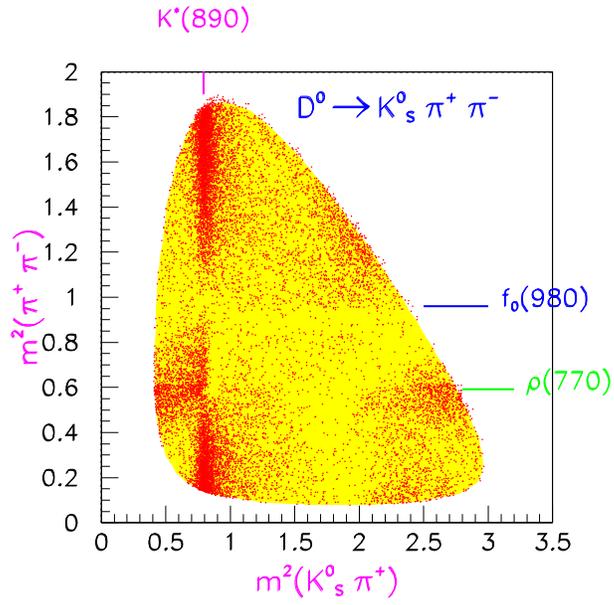}
\caption{$K^0_S \pi^+ \pi^-$ Dalitz plot with no background 
subtraction.}
\label{fig:fig5}
\end{center}
\end{figure}
\begin{figure}[htp]
\begin{center}
\includegraphics[width=12.0cm,height=10.0cm]{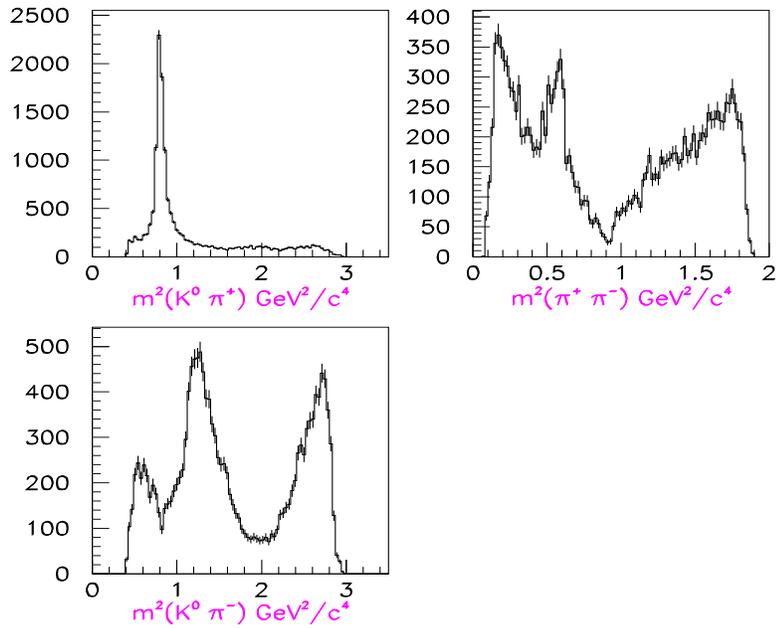}
\caption{$K^0_S \pi^+ \pi^-$ Dalitz plot projections.}
\label{fig:fig6}
\end{center}
\end{figure}
This Dalitz plot reveals a complex structure. 
Several intermediate resonant states involving $K^{*+}(892)$, $K^{*+}_0(1430)$,
$\rho^0(770)$, $f_0(980)$, $f_0(1370)$ resonances can be seen. Strong 
interferences can also be seen. In particular, $f_0(980)$ resonance shows up
as a uniform horizontal depletion suggesting interference with a broad scalar
resonance. 

\section{Selection of $D^0 \to K^0_S K^{\pm} \pi^{\mp}$}

The decay $D^0 \to K^0_S K \pi$ contains two possible
$D^0$ decay modes:
$$ D^0 \to K^0 K^- \pi^+ \qquad (a)$$
$$ D^0 \to \bar K^0 K^+ \pi^- \qquad (b)$$
The charge of the pion separates the two decay modes. Diagrams contributing to
the two decay channels are shown in fig.~\ref{fig:fig7}.
\begin{figure}[htp]
\begin{center}
\includegraphics[width=10.0cm,height=10.0cm]{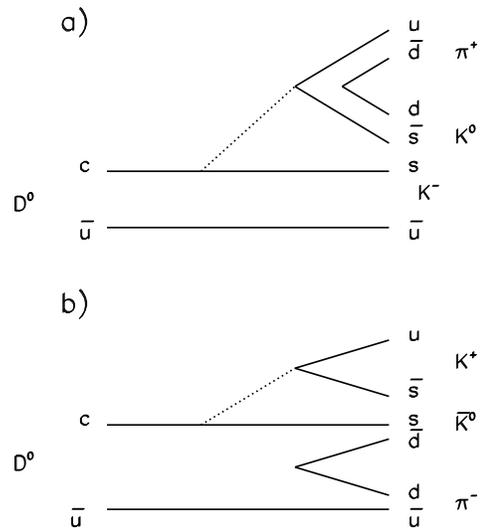}
\caption{Diagrams which contribute to the two different $D^0$ decay modes
to $K^0_S K \pi$.}
\label{fig:fig7}
\end{center}
\end{figure}
\begin{figure}[htp]
\begin{center}
\includegraphics[width=11.0cm,height=7.0cm]{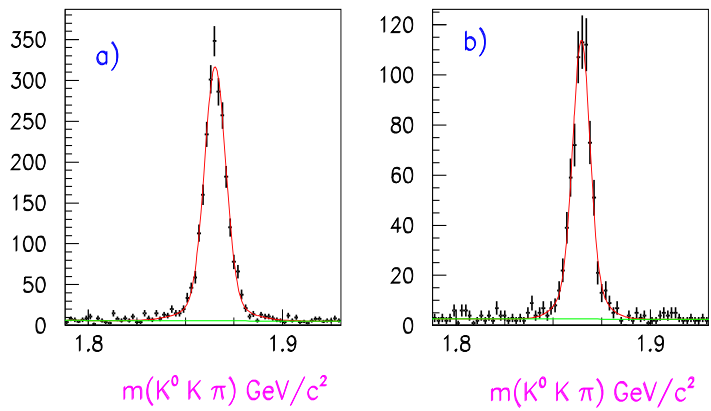}
\caption{a) $D^0 \to K^0 K^- \pi^+$ and b) $D^0 \to \bar K^0 K^+ \pi^-$ signals.}
\label{fig:fig8}
\end{center}
\end{figure}
\begin{figure}[htp]
\begin{center}
\begin{tabular}{cc}
\includegraphics[width=9.0cm,height=9.0cm]{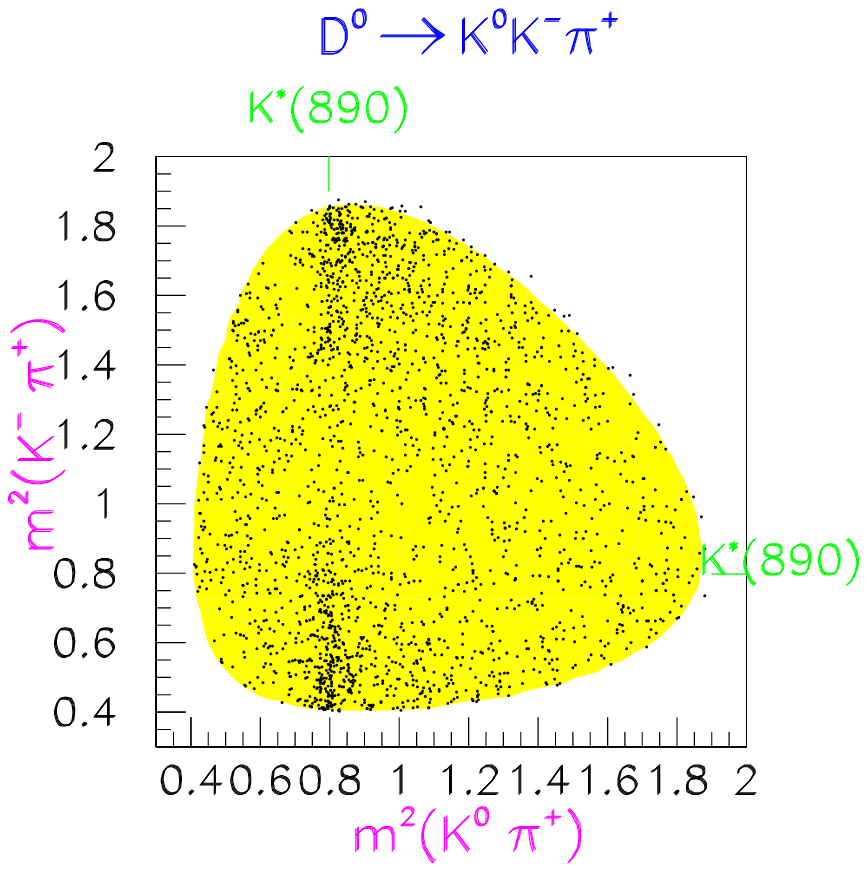} & \includegraphics[width=9.0cm,height=9.0cm]{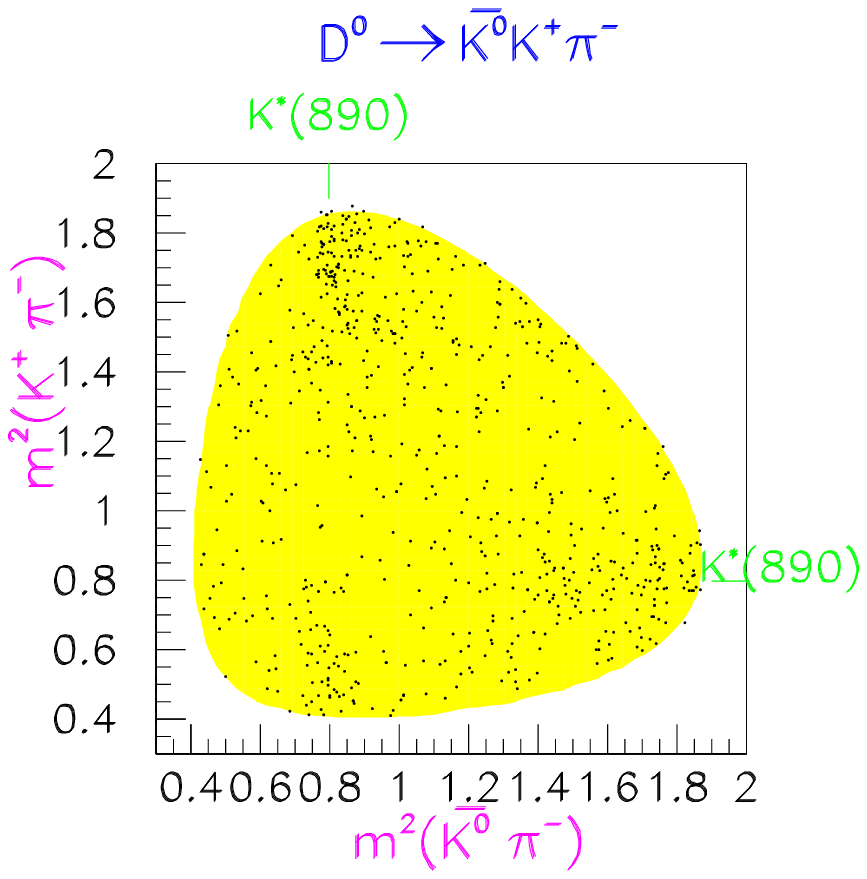}
\end{tabular}
\caption{a) $D^0 \to K^0 K^- \pi^+$ and b) $D^0 \to \bar K^0 K^+ \pi^-$ Dalitz
plots with no background subtraction.}
\label{fig:fig9}
\end{center}
\end{figure}
\begin{figure}[htp]
\begin{center}
\includegraphics[width=12.0cm,height=10.0cm]{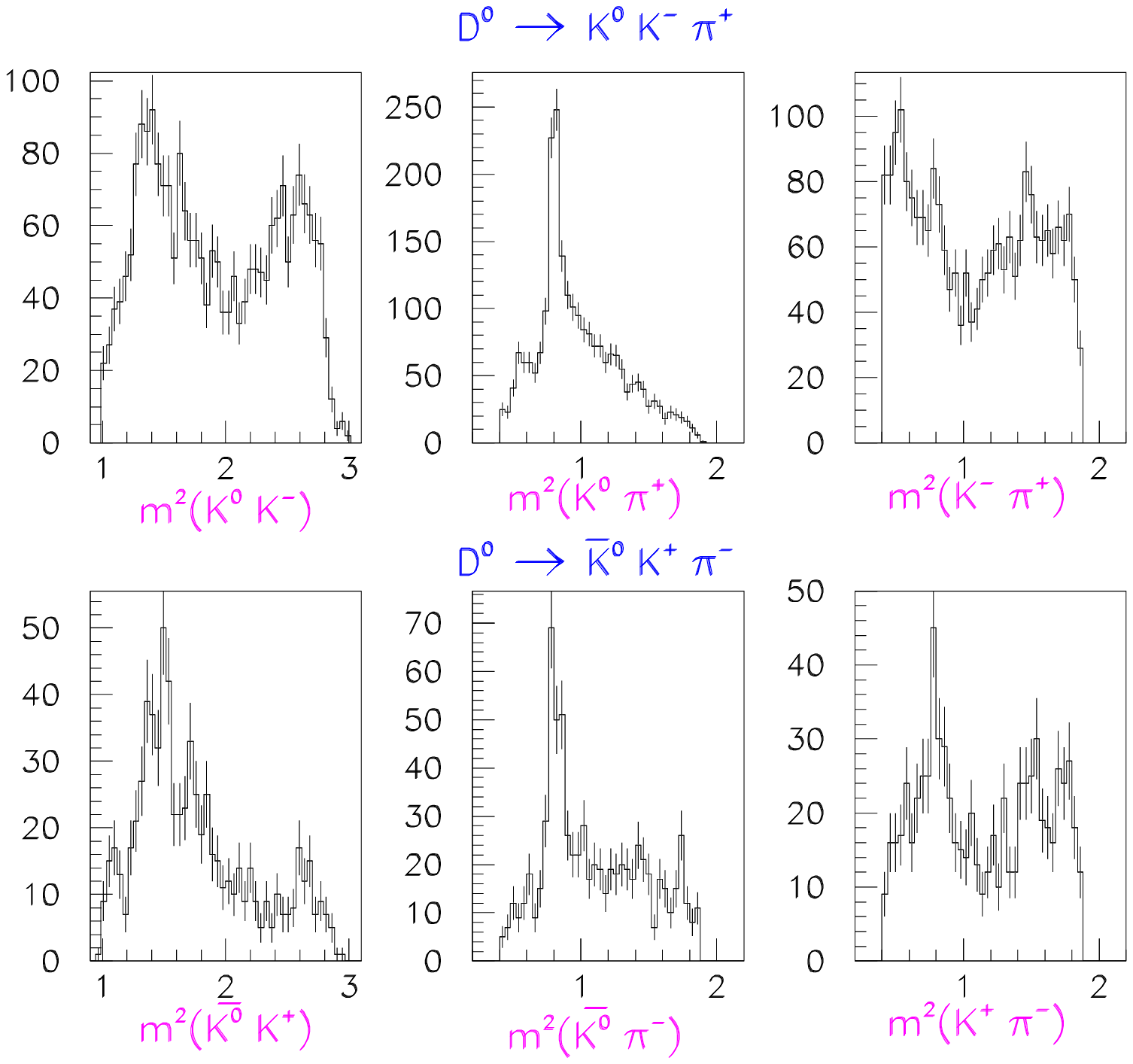}
\caption{First row: $D^0 \to K^0 K^- \pi^+$ Dalitz plot projections. Second
row: $D^0 \to \bar K^0 K^+ \pi^-$ Dalitz plot projections.}
\label{fig:fig10}
\end{center}
\end{figure}

The channels have been isolated using similar cuts on the corresponding 
$\Delta m$ distribution. Requiring 
one of the two charged tracks to be positively identified as a kaon we obtain the
$K^0_S K \pi$ mass distributions for the two decay modes shown in fig.~\ref{fig:fig8}.

The fitted  widths  and yields obtained using only one Gaussian and a linear 
background are the 
following:
$$(a) \quad D^0 \to K^0 K^- \pi^+: \quad (6.0 \pm 0.1) \quad MeV/c^2 \quad 2335 \quad \mathrm{events}$$
$$(b) \quad D^0 \to \bar K^0 K^+ \pi^-: \quad (5.1 \pm 0.1) \quad MeV/c^2 \quad 731 \quad \mathrm{events}$$
The above yields indicate that branching fractions for the above $D^0$ decay
channels are different.

The two Dalitz plots are shown in fig.~\ref{fig:fig9} and their projections
are shown in fig.~\ref{fig:fig10}.
The background fractions for decays a) and b) have been estimated to 
be 4 \% and 5\% respectively and have not been subtracted.

The decay $D^0 \to K^0 K^- \pi^+$ is dominated by $K^{*0} K^-$ with a 
small contribution from $\bar K^{*0} K^0$. We also observe the presence
of broad structure.
The decay $D^0 \to \bar K^0 K^+ \pi^-$ shows more symmetric $K^{*-} K^+$ and $K^{*0} \bar K^0$ contributions. 
\begin{figure}[htp]
\begin{center}
\includegraphics[width=7.0cm,height=7.0cm]{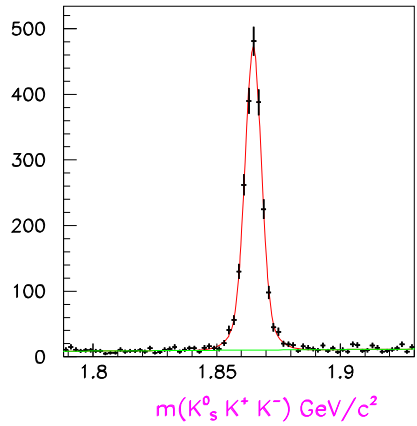}
\caption{$K^0_S K^+ K^-$ mass spectrum}
\label{fig:fig11}
\end{center}
\end{figure}
\begin{figure}[htp]
\begin{center}
\includegraphics[width=9.0cm,height=9.0cm]{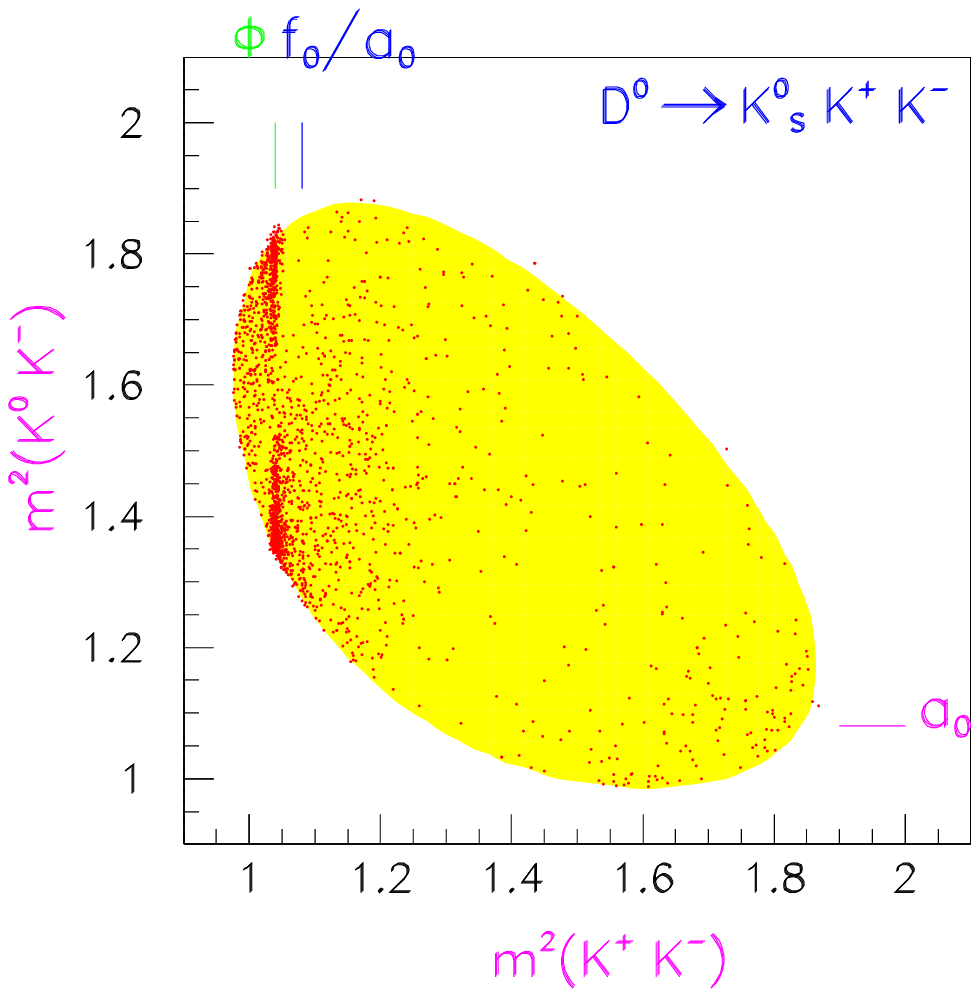}
\caption{$D^0 \to \bar K^0 K^+ K^-$ Dalitz plot.}
\label{fig:fig12}
\end{center}
\end{figure}
\begin{figure}[htp]
\begin{center}
\includegraphics[width=12.0cm,height=10.0cm]{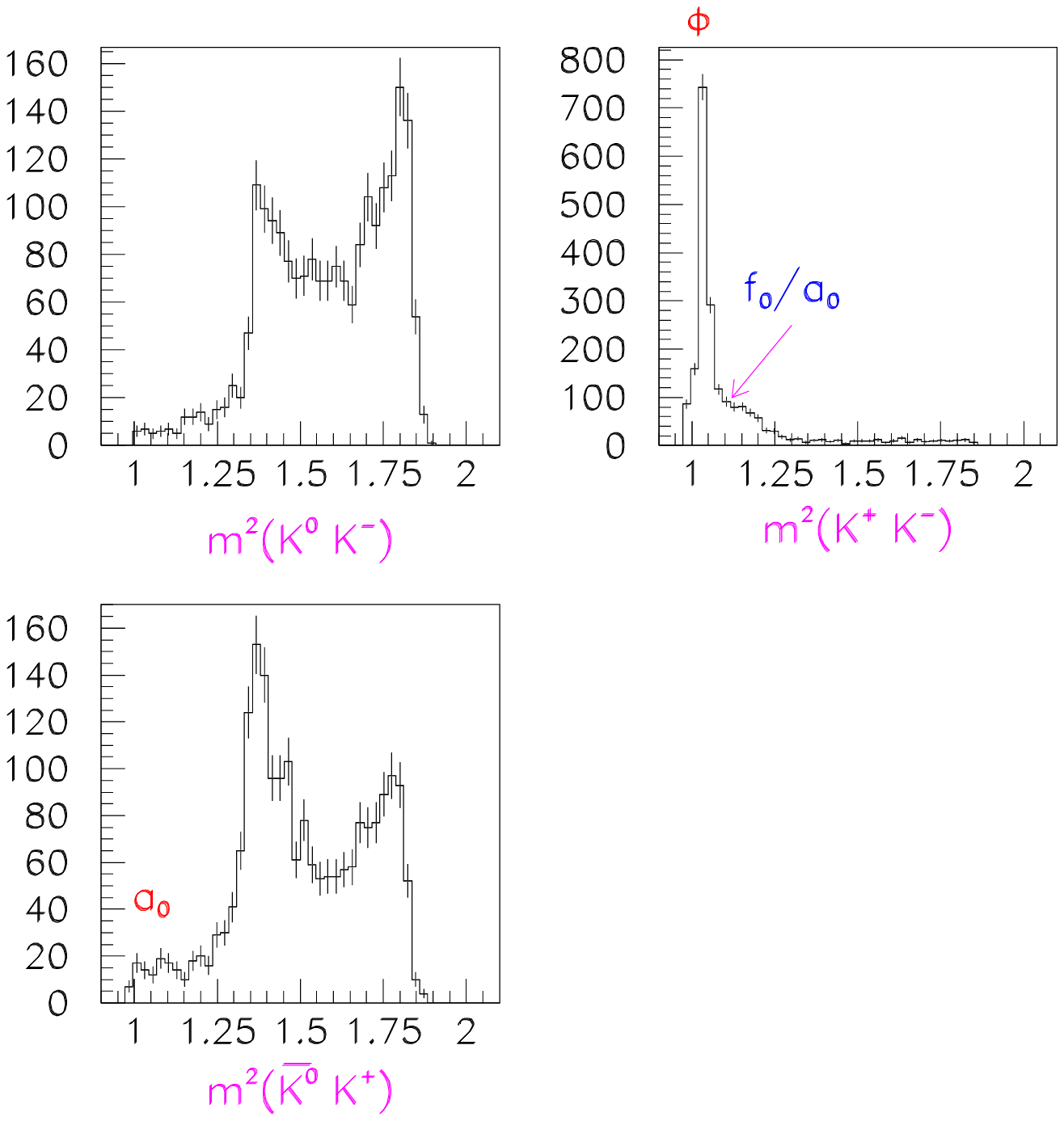}
\caption{$D^0 \to \bar K^0 K^+ K^-$ Dalitz plot projections.}
\label{fig:fig13}
\end{center}
\end{figure}

\section{Selection of $D^0 \to K^0_S K^+ K^-$}

The channel has been isolated performing a similar $\Delta m$ cut and 
requiring at least  
one of the two charged tracks to be positively identified as a kaon. 
The $K^0_S K^+ K^-$ mass spectrum is shown in fig.~\ref{fig:fig11}.

The $D^0$ width obtained fitting the mass spectrum with one Gaussian and a linear
background is (3.7 $\pm$ 0.1) $MeV/c^2$ and the 
yield is 2089 events.

The Dalitz plot of $D^0 \to K^0_S K^+ K^-$ is shown in fig.~\ref{fig:fig12} and 
contains a background fraction of 3 \% (not subtracted).
The corresponding projections are shown in fig.~\ref{fig:fig13}.

The presence of intermediate states involving $\phi$, $f_0(980)$ and 
$a_0(980)$ resonances is clearly visible.

\section{Conclusions}

Charm Physics can be performed at B-factories with high
statistics and small backgrounds. The Dalitz plot analysis as well as 
measurements of CP asymmetries in the decay amplitudes and rates 
for different charmed mesons are in progress.

In the near future, Charm Physics will be dominated by B-factories and $\tau$/charm 
factories. Present available statistics for Dalitz charm decays from 
fixed target and B-factories are: 1--5$\times 10^4$ events for Cabibbo allowed,
1--10$\times 10^3$ events for Cabibbo suppressed and 50--300 events for double
Cabibbo suppressed decays.

Given the large data samples being accumulated, we expect in the next few 
years an increase of these yields by a factor 20. This will allow charm
physics to be rewritten with errors on branching fractions reduced by more 
than a factor 10.

% choose bibtex style depending on layout style and options used in
% sample:

\end{document}